\documentclass[journal]{IEEEtran}                         

\IEEEoverridecommandlockouts                              

\usepackage{amsfonts}
\usepackage{mathrsfs}
\usepackage{amssymb}
\usepackage{amsmath}
\usepackage{array}
\usepackage{extarrows}
\usepackage{color}
\usepackage{cite}
\usepackage{graphicx}
\usepackage[font=footnotesize]{caption}
\usepackage[font=footnotesize]{subcaption}
\usepackage{stfloats}
\usepackage{epstopdf}
\usepackage{arydshln}
\usepackage{ifthen}
\usepackage{url}
\usepackage{pict2e}
\DeclareMathOperator{\g-rank}{grk \,}

\makeatletter 
\newsavebox\myboxA
\newsavebox\myboxB
\newlength\mylenA
\newcommand*\xoverline[2][0.75]{%
    \sbox{\myboxA}{$\m@th#2$}%
    \setbox\myboxB\null
    \ht\myboxB=\ht\myboxA%
    \dp\myboxB=\dp\myboxA%
    \wd\myboxB=#1\wd\myboxA
    \sbox\myboxB{$\m@th\overline{\copy\myboxB}$}
    \setlength\mylenA{\the\wd\myboxA}
    \addtolength\mylenA{-\the\wd\myboxB}%
    \ifdim\wd\myboxB<\wd\myboxA%
       \rlap{\hskip 0.5\mylenA\usebox\myboxB}{\usebox\myboxA}%
    \else
        \hskip -0.5\mylenA\rlap{\usebox\myboxA}{\hskip 0.5\mylenA\usebox\myboxB}%
    \fi}

\input{steve.lib}

\title{\LARGE \bf
A Graphical Characterization of Structurally Controllable Linear Systems with Dependent Parameters
}

\author{F. Liu and A. S. Morse
\thanks{This work was supported by National Science Foundation grant n. 1607101.00 and US Air Force grant n. FA9550-16-1-0290.}
\thanks{F. Liu and A. S. Morse are with the Department
of Electrical Engineering, Yale University, New Haven, CT 06511 USA (e-mail: fengjiao.liu@yale.edu; as.morse@yale.edu).}%
\thanks{A short version \cite{liu2017structural} of this paper which omitted proofs of all lemmas was presented at the $56$th IEEE Conference on Decision and Control in December, 2017.}%
        }

\begin{document}
\maketitle
\thispagestyle{empty} 
\pagestyle{empty}

\begin{abstract}
One version of the concept of structural controllability defined for single-input systems by Lin and subsequently generalized to multi-input systems by others, states that a parameterized matrix pair $(A, B)$ whose nonzero entries are distinct parameters, is structurally controllable if values can be assigned to the parameters which cause the resulting matrix pair to be controllable. In this paper the concept of structural controllability is broadened to allow for the possibility that a parameter may appear in more than one location in the pair $(A, B)$. Subject to a certain condition on the parameterization called the ``binary assumption", an explicit graph-theoretic characterization of such matrix pairs is derived.
\end{abstract}


\section{Introduction}

Over the past few years there has been a resurgence of interest in the question of structural controllability posed by Lin in 1974 \cite{lin1974structural}, which aims to capture the controllability of systems with parameters whose values are not exactly known but only approximately determined. As defined by Lin, a pair of matrices $(A_{n \times n}, b_{n \times 1})$ with each entry either a fixed zero or a distinct scalar parameter is \emph{structurally controllable} if there is a real matrix pair $(\bar{A}_{n \times n}, \bar{b}_{n \times 1})$ with the same pattern of zero entries as $(A, b)$ which is controllable. Thus if $(A, b)$ is structurally controllable, then almost every real matrix pair $(\bar{A}, \bar{b})$ with the same pattern of zero entries as $(A, b)$ will be controllable. Lin was able to give an explicit graph-theoretic condition for such a matrix pair to be structurally controllable in terms of properties of a suitably defined directed graph determined by the given matrix pair. Lin's result was extended to multi-input matrix pairs $(A_{n \times n}, B_{n \times m})$ in linear algebra terms by Shields and Pearson \cite{shields1975structural} and reexplained in graph theory terms by Mayeda \cite{mayeda1981structural}. Generic properties and design problems of Lin's parameterization of the pair $(A, B)$ were studied in \cite{lin1977system, hosoe1980determination, dion2003generic, maza2012impact, zufiria2014mathematical, olshevsky2015minimum, pequito2016framework, zhang2017edge}. Results on structural controllability of linear time-varying systems were presented in \cite{poljak1992gap, hartung2013necessary, liu2013structural, pan2014structural, posfai2014structural, garcia2015structural, hou2016structural, pequito2017structural, yao2017structural}. One line of research deals with the structural controllability of composite systems \cite{rech1991structural, li1996g, blackhall2010structural, carvalho2017composability, mu2018structural}. Current interest stems from the realization that structural controllability is a key property of interest in swarming behavior and in the modeling and understanding complex networks \cite{liu2011controllability, cowan2012nodal, nacher2013structural, zhang2014structural, yin2016minimum, sun2016towards, tang2017developmental, riasi2017controllability, guan2017structural}. For example, identification, characterization, and classification of driver vertices or steering vertices in biomedical networks \cite{liu2014detection, liu2015identifying, wu2016minimum, chu2017wdnfinder, guo2017constrained, yao2017functional, kanhaiya2017controlling, ravindran2017identification}, which tend to have strong ability to influence other vertices, may enlighten us on critical underlying relations or mechanisms; the study of robustness of structural controllability to vertex and/or arc failures and disturbances \cite{rech1990robustness, rahimian2013structural, wang2013maintain, rahimian2013links, mengiste2015effect, zhang2018optimization} may give us an insight in the evolution of complex social networks and various issues of network security \cite{pasqualetti2015control, weerakkody2017graph}.

In previous work, there was also interest in structural controllability for more general kinds of parameterizations \cite{corfmat1976structurally, hayakawa1984structural, willems1986structural, anderson1982structural}. In particular, the notion of a ``linearly parameterized" matrix pair was introduced in \cite{corfmat1976structurally} which allowed parameters to appear in multiple locations in the pair $(A, B)$. The engineering motivation for linear parameterization came from physical systems with unknown but dependent design parameters involving the imprecise values of their physical components \cite{dasgupta1987physically} or measurements \cite{hayakawa1984structural}. 
\begin{figure}[!h]
    \centering
    \includegraphics[width=0.147\textwidth]{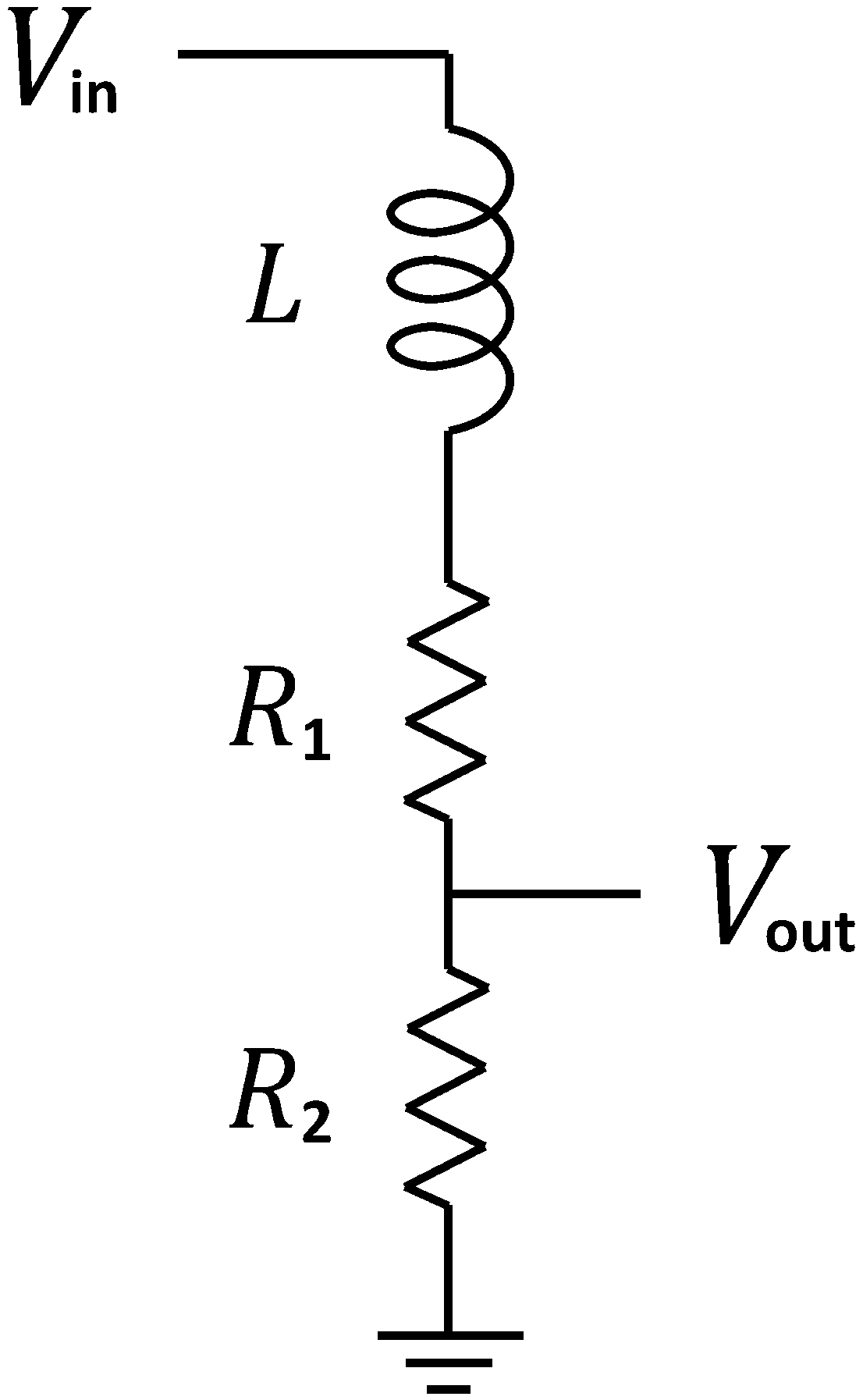}
    \caption{A voltage divider circuit.}
    \label{fig:circuit}
\end{figure}

For example, the transfer function of the voltage divider circuit in Figure~\ref{fig:circuit} is
\begin{equation*}
T(s) = \frac{V_{\text{out}}(s)}{V_{\text{in}}(s)} = \frac{R_2}{R_1 + R_2 + sL}
\end{equation*}
A one-dimensional realization of the transfer function is 
\begin{equation*}
\dot{x} = - \frac{R_1 + R_2}{L} \enspace x + \frac{R_2}{L} \enspace V_{\text{in}}, \qquad V_{\text{out}} = x
\end{equation*}
So for this system, $A = - \frac{R_1}{L} - \frac{R_2}{L}$ and $B = \frac{R_2}{L}$. Note that $\frac{R_2}{L}$ appears in both $A$ and $B$. Suppose the exact values of the physical components $R_1$, $R_2$, and $L$ are unknown, let $p_1 = \frac{R_1}{L}$ and $p_2 = \frac{R_2}{L}$, then $A$ and $B$ are linear functions of the two independent parameters $p_1$ and $p_2$.

In this paper, we address this same kind of parameterization satisfying a certain condition called the ``binary assumption" and show by counterexample that this is the most general class of linear parameterizations for which one can expect a graphical characterization with unweighted graphs. Finally, the structural controllability of this class of linear parameterizations is characterized in strict graph-theoretic terms, which provides a guide to designing and analyzing complex networks with coupled links.

\subsection{Linear Parameterization}

Interesting as the results of Lin's parameterization are, they cannot address many simple but commonly encountered modeling situations such as when $A$ and $b$ are of the forms
\begin{equation} \label{eqn:lnr-prm-exmp}
A =
\begin{bmatrix}
p_1 & p_1 \\
0 & 0
\end{bmatrix}
, \quad
b =
\begin{bmatrix}
p_1 \\
p_2
\end{bmatrix}
\end{equation}
where at least one parameter, in this example $p_1$, appears in more than one location. Recognition of this led to the definition of a ``linearly parameterized" matrix pair and to a significant generalization of the concept of structural controllability \cite{corfmat1976structurally}. The version of a \emph{linearly parameterized} matrix pair to which we are referring is of the form\footnote{Although written differently, this linear parameterization is in fact the same as the one considered in \cite{corfmat1976structurally} except that in \cite{corfmat1976structurally} there are constant matrices $A_0$ and $B_0$ also appearing in the sums in (\ref{eqn:lnr-prm-pair-sep}) for $A(p)$ and $B(p)$ respectively.}
\begin{equation} \label{eqn:lnr-prm-pair-sep}
A_{n \times n}(p) = \sum_{k \in \mathbf{q}} g_k p_k h_{k1}, ~~
B_{n \times m}(p) = \sum_{k \in \mathbf{q}} g_k p_k h_{k2} 
\end{equation} 
where $p \in {\rm I\!R}^q$ is a vector of $q > 0$ algebraically independent parameters $p_1$, $p_2$, $\dots$, $p_q$, $\mathbf{q} \triangleq \{1, 2, \dots, q\}$, and for each $k \in \mathbf{q}$, $g_k \in {\rm I\!R}^n$, $h_{k1} \in {\rm I\!R}^{1 \times n}$, $h_{k2} \in {\rm I\!R}^{1 \times m}$. In this context, the problem of interest is to find conditions for the existence of a parameter vector $p \in {\rm I\!R}^q$ for which $(A(p),B(p))$ is a controllable matrix pair. If such values exist, the parameterized pair $(A(p),B(p))$ is \emph{structurally controllable}. Such pairs are controllable for almost every value of $p$ in the sense that the set of values of $p$ for which $(A, B)$ is controllable is the complement of a proper algebraic set in ${\rm I\!R}^q$.

Necessary and sufficient conditions for such a matrix pair to be structurally controllable in this more generalized sense are developed in \cite{corfmat1976structurally}. Like the work of Shields and Pearson \cite{shields1975structural}, these conditions are primarily matrix-algebraic. A special form of linearly parameterized matrix pairs can be used to model compartmental systems and corresponds to compartmental graphs, on which graphical conditions for the structural controllability of matrix pairs in this special form have been investigated \cite{hayakawa1984structural}. Other types of parameterization have also been explored, but either purely algebraically \cite{willems1986structural, anderson1982structural}, or without equivalent graphical conditions \cite{mousavi2018structural}. Since graphical results can reveal important structural properties hidden in matrix representations, there is interest in determining graphical conditions characterizing the generalized concept of structural controllability and this is the specific problem which this paper is addressed.

Before proceeding we point out that not every matrix pair $(A, B)$ with parameters entering ``linearly" is a linear parameterization as defined here. For example, while the matrix pair shown in (\ref{eqn:lnr-prm-exmp}) is linearly parameterized, the matrix pair
\begin{equation} \label{eqn:lnr-prm-cntexmp}
A =
\begin{bmatrix}
p_1 & p_1 \\
0 & p_1
\end{bmatrix}
, \quad
b =
\begin{bmatrix}
0 \\
p_2
\end{bmatrix}
\end{equation}
is not. It is claimed that a matrix pair $(A, B)$ whose entries depend linearly on $q$ parameters $p_1$, $p_2$, $\dots$, $p_q$ will be linearly parameterized if and only if all minors of the partitioned matrix $[A \enspace B]$ are multilinear functions of the $q$ parameters. It is clear that the matrices in (\ref{eqn:lnr-prm-cntexmp}) do not have this property. To see why the claim is true, let $(A, B)$ be a linearly parameterized matrix pair and fix the values of all parameters except for $p_k$. If $p_k$ appears in only one row or column of a square submatrix of $[A \enspace B]$, its determinant is a linear function of $p_k$. If $p_k$ appears in more rows and columns of a square submatrix, it must enters the matrix in a rank-one fashion, as $\rank (g_k \, [\, h_{k1} \enspace h_{k2} \,]) = 1$. So by adding scalar multiples of one row that contains $p_k$ to other rows containing $p_k$, it is possible to get another square matrix of the same determinant, with $p_k$ appearing in only one row. Therefore all minors of $[A \enspace B]$ are multilinear functions of the $q$ parameters. The statement in other direction can be easily proved by its contrapositive.

In the sequel it will be convenient to use the $n\times (n+m)$ partitioned matrix $[A(p) \enspace B(p)]$. In view of (\ref{eqn:lnr-prm-pair-sep}), this matrix can be expressed as
\begin{equation} \label{eqn:lnr-prm-pair-cmb}
[A(p) \enspace B(p)] = \sum_{k \in \mathbf{q}} g_k p_k h_k
\end{equation}
where $h_k \triangleq [h_{k1} \enspace h_{k2}]$ for $k \in \mathbf{q}$. It will be assumed for simplicity and without loss of generality that the set of matrices $\{g_1 h_1, g_2 h_2, \dots, g_q h_q\}$ is linearly independent. To justify this assumption, suppose that the set is not linearly independent and for purposes of illustration that $g_qh_q$ is a linear combination of the remaining matrices $g_1 h_1, g_2 h_2, \dots, g_{(q-1)} h_{(q-1)}$. In other words, suppose that
\begin{equation*}
g_q h_q = \sum_{k = 1}^{q-1} c_k g_k h_k
\end{equation*}
where the $c_k$ are real numbers. Then in view of (\ref{eqn:lnr-prm-pair-cmb}),
\begin{equation*}
[A(p) \enspace B(p)] = \sum_{k = 1}^{q-1} g_k (p_k + c_k p_q) h_k
\end{equation*}
Therefore if we define new parameters $\bar{p}_k = p_k + c_k p_q$ for $k \in \{1, 2, \dots, q-1\}$, then the right side of (\ref{eqn:lnr-prm-pair-cmb}) can be written using only the first $q-1$ matrices in $\{g_1 h_1, g_2 h_2, \dots, g_q h_q\}$ as 
\begin{equation*}
\sum_{k = 1}^{q-1} g_k \bar{p}_k h_k
\end{equation*}
It is clear from this that the process of defining new parameters and eliminating dependent matrices from $\{g_1 h_1, g_2 h_2, \dots, g_q h_q\}$ can be continued until one has a linearly independent subset of matrices. This justifies our claim and accordingly it will henceforth be assumed that $\{g_1 h_1, g_2 h_2, \dots, g_q h_q\}$ is a linearly independent set. This implies that $q \leq n(n+m)$.

Since this paper deals with matrix pairs parameterized by $p$ exclusively, it is convenient to drop $p$ in $A(p)$ and $B(p)$, i.e., to write $A$ and $B$ instead of $A(p)$ and $B(p)$, with the understanding that $(A, B)$ is parameterized by $p$.

\subsection{Graph of $(A, B)$}

It is easy to see that the definition of structural controllability for a linearly parameterized matrix pair $(A, B)$ coincides with Lin's if $m=1$ and the $g_k$ and $h_k$ are restricted to be unit vectors in ${\rm I\!R}^n$ and ${\rm I\!R}^{1 \times (n+1)}$ respectively. Lin defines the graph of such a matrix pair to be an unweighted directed graph on $n+1$ vertices labeled $1$ to $n+1$ with an arc from vertex $j$ to vertex $i$ if the $ij$th entry in the matrix $[A \enspace B]$ is a parameter. For the more general linear parameterization defined by (\ref{eqn:lnr-prm-pair-cmb}), a more elaborate definition of a graph is needed not just because $m$ might be greater than 1, but also because some parameter $p_k$ may appear in multiple locations in $[A \enspace B]$.

The \emph{graph} of $(A, B)$, written $\mathbb{G}$, is defined to be an unweighted directed graph with $n+m$ vertices labeled $1$ through $n+m$ and an arc of color\footnote{In this paper, each color is labeled by a distinct integer.} $k$ from vertex $j$ to vertex $i$ if the $ij$th entry in the matrix $g_k h_k$ is nonzero, i.e., the $ij$th entry in the partitioned matrix $[A \enspace B]$ contains $p_k$. In the sequel, $(j, i)_k$ denotes an arc from vertex $j$ to vertex $i$ with color $k$. This graph has $q$ colors.

\begin{figure}[!h]
    \centering
    \includegraphics[width=0.387\textwidth]{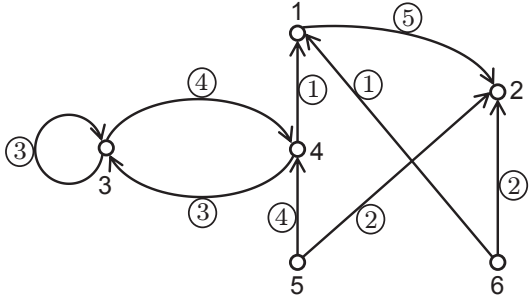}
    \caption{The graph of the matrix pair in (\ref{eqn:bin-prm-exmp}).}
    \label{fig:graph-example}
\end{figure}

Figure~\ref{fig:graph-example} shows the graph of 
\begin{equation} \label{eqn:bin-prm-exmp}
\left(
\begin{bmatrix}
0 & 0 & 0 & p_1 \\
p_5 & 0 & 0 & 0 \\
0 & 0 & p_3 & p_3 \\
0 & 0 & p_4 & 0 
\end{bmatrix}
, 
\begin{bmatrix}
0 & p_1 \\
p_2 & p_2 \\
0 & 0 \\
p_4 & 0
\end{bmatrix}
\right)
\end{equation}
where symbol \textcircled{\footnotesize{$k$}} labels color $k$ for $k = 1, 2, \dots, 5$.

Note that the graph of $(A, B)$ has three properties: (\lowercase\expandafter{\romannumeral1}) There is no arc pointing toward any of the $m$ vertices with labels $n+1$ to $n+m$, since the matrix $[A \enspace B]$ only has $n$ rows. (\lowercase\expandafter{\romannumeral2}) There may be more than one arc from one given vertex $j$ to another vertex $i$, for the $ij$th entry in the matrix $[A \enspace B]$ may be a linear combination of more than one parameter. If this is the case, all arcs from vertex $j$ to vertex $i$ will have distinct colors. (\lowercase\expandafter{\romannumeral3}) If there are two arcs of color $k \in \mathbf{q}$, one leaving vertex $j$ and the other pointing toward vertex $i$, then there must be an arc $(j, i)_k$. This is because the two given arcs imply that the $j$th entry in the row vector $h_k$ and the $i$th entry in the column vector $g_k$ are nonzero, which means the $ij$th entry in the matrix $g_k h_k$ is nonzero. Any unweighted directed graph on $n+m$ vertices which has these properties is called a \emph{structural controllability graph}.

\subsection{Binary Assumption}

This paper focuses exclusively on linear parameterizations which satisfy a certain ``binary assumption". Specifically, the linear parameterization defined by (\ref{eqn:lnr-prm-pair-cmb}) satisfies the \emph{binary assumption} if all of the $g_k$ and $h_k$ appearing in (\ref{eqn:lnr-prm-pair-cmb}) are binary vectors, i.e., vectors of $1$'s and $0$'s. Similarly, a linear parameterization satisfies the \emph{unitary assumption} if all of the $g_k$ and $h_k$ appearing in (\ref{eqn:lnr-prm-pair-cmb}) are unit vectors. So any linear parameterization satisfying the unitary assumption also satisfies the binary assumption. Lin's parameterization is exactly the linear parameterization satisfying the unitary assumption.

It is quite clear that when the binary assumption holds with $n$ and $m$ specified, the parameterization in (\ref{eqn:lnr-prm-pair-cmb}) is uniquely determined by a structural controllability graph. Because of this, it is possible to characterize the structural controllability of a linearly parameterized matrix pair $(A,B)$ which satisfies the binary assumption, solely in terms of the graph of the pair. On the other hand, without the binary assumption, no such graphical characterization\footnote{If the binary assumption is dropped, one way to proceed is to define the graph of $(A, B)$ as a weighted directed graph, in which the weight of an arc $(j, i)_k$ is the $ij$th entry in the matrix $g_k h_k$. The conditions on weighted graphs for the structural controllability of all linearly parameterized matrix pairs will be studied in a sequel of this paper.} is possible. The following example illustrates this.

Note that although the matrix pairs
\begin{equation*}
\left(
\begin{bmatrix}
p_1 & p_1 & p_2 \\
p_1 & p_1 & p_2 \\
0   & 0   & 0
\end{bmatrix}
, 
\begin{bmatrix}
0 \\
0 \\
p_3
\end{bmatrix} 
\right)
~ \text{and} ~
\left(
\begin{bmatrix}
p_1 & p_1 & p_2 \\
p_1 & p_1 & 2p_2 \\
0   & 0   & 0
\end{bmatrix}
, 
\begin{bmatrix}
0 \\
0 \\
p_3
\end{bmatrix}
\right)
\end{equation*}
both have the same graph, only the pair on the right is structurally controllable. Of course the pair on the right does not satisfy the binary assumption.

\subsection{Problem Formulation and Organization}

This paper gives necessary and sufficient graph-theoretic conditions for the structural controllability of a linearly parameterized matrix pair $(A_{n \times n}, B_{n \times m})$ which satisfies the binary assumption. To the best of our knowledge, this is the first graph-theoretic result that generalizes the conditions given in \cite{lin1974structural} and \cite{mayeda1981structural}. The rest of the paper is organized as follows. The terminology and concepts used in this paper are defined in Section \uppercase\expandafter{\romannumeral2}. The main result of this paper is presented in Section \uppercase\expandafter{\romannumeral3} and proved in Section \uppercase\expandafter{\romannumeral4}.

\section{Preliminaries}

In order to state the main result of this paper, some terminology and a number of graphical and algebraic concepts are needed. 
\begin{figure*}[!h]
	\centering
	\begin{subfigure}{0.168\textwidth}
		\includegraphics[width=\linewidth]{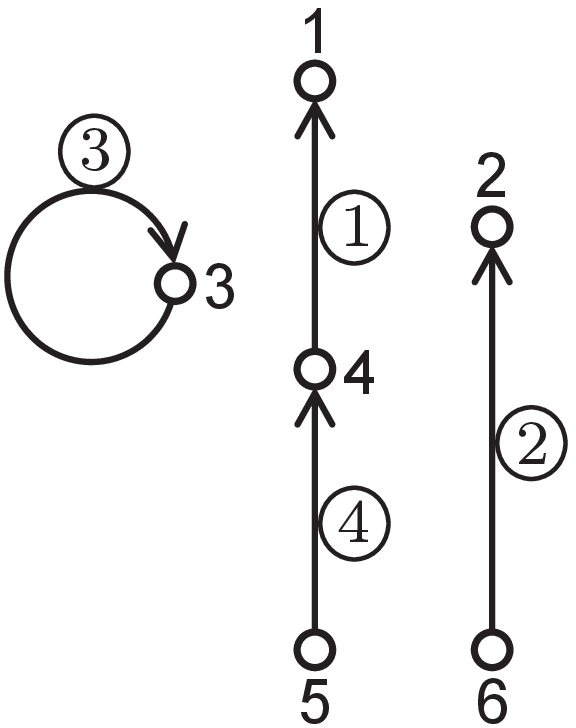}
		\caption{} 
		\label{subfig:odd-rep}
	\end{subfigure}
	\hfil  \hfil  \hfil  \hfil  \hfil  \hfil
	\begin{subfigure}{0.224\textwidth}
		\includegraphics[width=\linewidth]{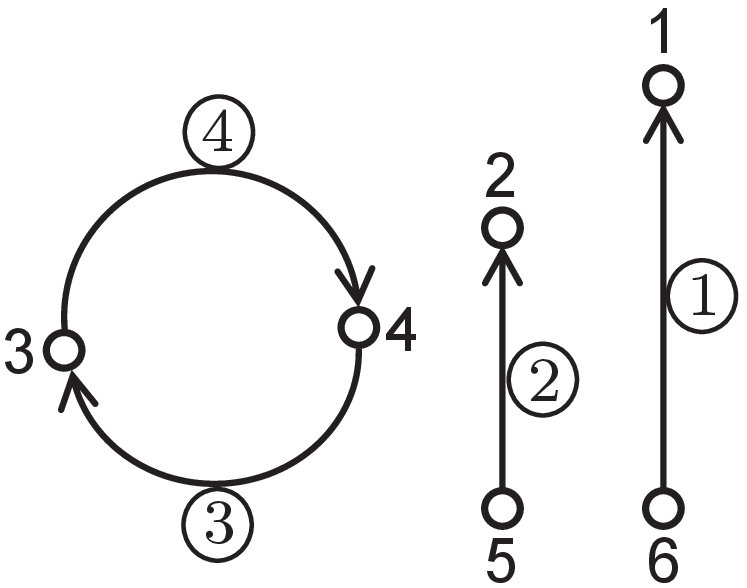}
		\caption{} 
		\label{subfig:even-rep}
	\end{subfigure}
	\hfil  \hfil  \hfil  \hfil  \hfil  \hfil
	\begin{subfigure}{0.159\textwidth}
		\includegraphics[width=\linewidth]{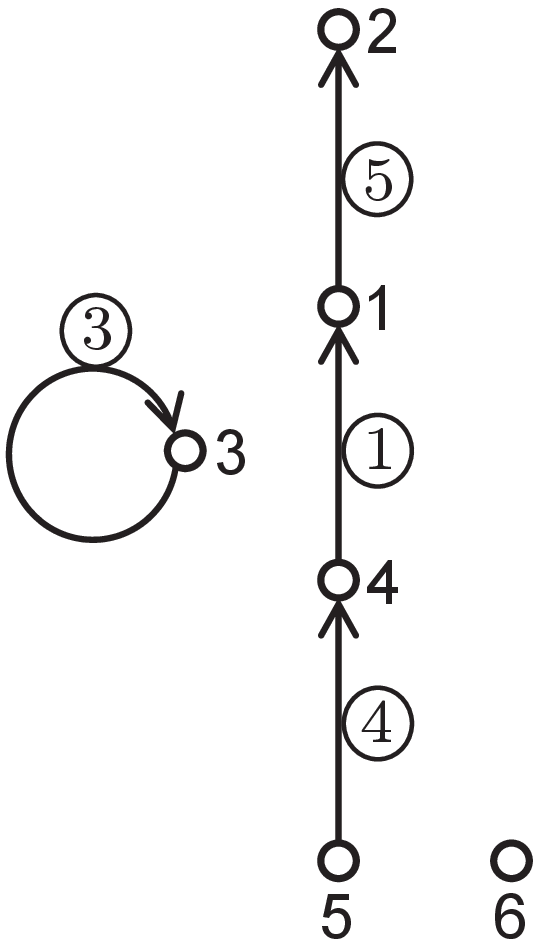}
		\caption{} 
		\label{subfig:single-rep}
	\end{subfigure}
	\caption{Multi-colored subgraphs of the graph in Figure~\ref{fig:graph-example}.} 
	\label{fig:odd-even-sgl-rep}
\end{figure*}

\subsection{Terminology}

Let $\mathbb{H}$ be an unweighted directed graph with a vertex set $\mathcal{V}$ and an arc set $\mathcal{A}$. An \emph{induced subgraph} of $\mathbb{H}$ by a subset of vertices $\mathcal{U} \subset \mathcal{V}$ is a subgraph of $\mathbb{H}$, whose vertex set is $\mathcal{U}$ and whose arc set is $\{(i, j) \, | \, i, j \in \mathcal{U}, (i, j) \in \mathcal{A}\}$. For any subset $\mathcal{U} \subset \mathcal{V}$, $\mathcal{V} - \mathcal{U}$ is the complement of $\mathcal{U}$ in $\mathcal{V}$. A \emph{source vertex} in $\mathbb{H}$ is a vertex with no incoming arc and a \emph{sink vertex} in $\mathbb{H}$ is a vertex with no outgoing arc. An \emph{isolated vertex} is both a source vertex and a sink vertex. A \emph{partition} $\pi$ of $\mathcal{V}$ is a family of nonempty subsets of $\mathcal{V}$ which are pairwise disjoint and whose union is equal to $\mathcal{V}$. The \emph{quotient graph} of $\mathbb{H}$ induced by $\pi$, written $\mathbb{H}/\pi$, is an unweighted directed graph with one vertex for each cell of $\pi$, and exactly one arc from vertex $i$ to vertex $j$ whenever $\mathbb{H}$ has at least one arc from the vertices in the $i$th cell to the vertices in the $j$th cell. $\mathbb{H}/\pi$ is the \emph{condensation} of $\mathbb{H}$ if $\pi$ is formed by the collection of strongly connected components of $\mathbb{H}$.

A \emph{directed path graph} is a weakly connected \cite{godsil2013algebraic} graph whose vertices can be labeled in the order $1$ to $k$ for some $k \in {\rm I\!N}$ such that the arcs are $(i, i+1)$, where $i=1, 2, \dots, k-1$. The \emph{length} of a directed path graph is the number of arcs in it. So in a directed path graph of positive length, the first vertex has exactly one outgoing arc, the last vertex has exactly one incoming arc, and each of the other vertices in between has exactly one incoming arc and one outgoing arc. In this context, a directed path graph of length $0$ is an isolated vertex. A \emph{directed cycle graph} is a strongly connected \cite{godsil2013algebraic} graph whose vertices can be labeled in the order $1$ to $k$ for some $k \in {\rm I\!N}$ such that the arcs are $(i, i+1)$ and $(k, 1)$, where $i=1, 2, \dots, k-1$. So in a directed cycle graph, each vertex has exactly one incoming arc and one outgoing arc. One vertex with a single self-loop is also a directed cycle graph. As this paper is concerned with directed graphs only, a directed path graph and a directed cycle graph will be simply called a path graph and a cycle graph, respectively, in the rest of the paper. The \emph{disjoint union} of two or more graphs is the union of these graphs whose vertex sets are disjoint. A directed graph is \emph{rooted} if it contains at least one vertex $r$ called a \emph{root} with the property that for each remaining vertex $v$ there is a directed path from $r$ to $v$. Rooted directed graphs arise naturally in the study of consensus problems \cite{cao2008reaching}. A \emph{directed rooted tree} is a rooted directed graph which is also a weakly connected tree \cite{godsil2013algebraic}. $\mathbb{H}$ has a \emph{spanning forest} if it has a spanning subgraph \cite{godsil2013algebraic} which is the disjoint union of directed rooted trees. Let $\mathcal{V}_{\text{root}} \subset \mathcal{V}$ be the set of root vertices of the trees. With a slight abuse of terminology, we will say that $\mathbb{H}$ has a spanning forest rooted at the vertices in $\mathcal{V}_{\text{root}}$ if and only if for each vertex $v \in \mathcal{V} - \mathcal{V}_{\text{root}}$, there is a path to $v$ from one of the root vertices.

\subsection{Graphical Concepts}

A \emph{multi-colored subgraph} of a structural controllability graph $\mathbb{G}$ is a spanning subgraph of $\mathbb{G}$, which is the disjoint union of $m$ path graphs and any number of cycle graphs with each arc in the union graph of a different color. Clearly, a multi-colored subgraph of $\mathbb{G}$ has $n$ arcs that do not share colors, start vertices, or end vertices. Figure~\ref{fig:odd-even-sgl-rep} shows three multi-colored subgraphs of the graph in Figure~\ref{fig:graph-example}.

A multi-colored subgraph $\mathbb{S}$ of a structural controllability graph $\mathbb{G}$ is obtained by sequentially removing arcs from $\mathbb{G}$ as follows. First pick any arc $(a, b)_{k_1}$ in $\mathbb{G}$ and then remove any other arcs with the same color $k_1$ as well as any arcs other than $(a, b)_{k_1}$ pointing toward vertex $b$ and/or leaving vertex $a$. Next, from the set of arcs which remain after these removals, pick any arc $(c, d)_{k_2}$ and repeat the process until no further arc picking is possible. If a total of $n$ arcs are left, the graph which remains is $\mathbb{S}$. Clearly, $\mathbb{S}$ is not unique. In the sequel, $\mathcal{R}(\mathbb{G})$ denotes the set of all multi-colored subgraphs of $\mathbb{G}$.

Suppose $\mathbb{G}$ is the graph of a linearly parameterized matrix pair $(A, B)$. It is possible that $\mathbb{G}$ does not have any multi-colored subgraph, that is, $\mathcal{R}(\mathbb{G})$ is an empty set. As the $n$ arcs in a multi-colored subgraph have $n$ distinct colors, $n$ different start vertices and $n$ different end vertices, $\mathcal{R}(\mathbb{G}) = \emptyset$ if and only if there are no $n$ distinct parameters appearing in $n$ different rows and $n$ different columns of the partitioned matrix $[A \enspace B]$. If so, $\rank [A \enspace B] < n$ for any $p \in {\rm I\!R}^q$, as each parameter enters $[A \enspace B]$ in a rank-one fashion. Then the pair $(A, B)$ is not structurally controllable.

The source vertices (respectively, sink vertices) of a multi-colored subgraph $\mathbb{S}$ are the $m$ source vertices (respectively, sink vertices) of the path graphs in $\mathbb{S}$. It is not hard to see that the source vertices of every multi-colored subgraph of $\mathbb{G}$ are the $m$ vertices with labels $n+1$ to $n+m$, since there is no arc pointing toward any of them. But the sink vertices of a multi-colored subgraph may be any $m$ vertices in $\mathbb{G}$. 
\begin{figure*}[!t]
	\centering
	\begin{subfigure}{0.332\textwidth}
		\includegraphics[width=\linewidth]{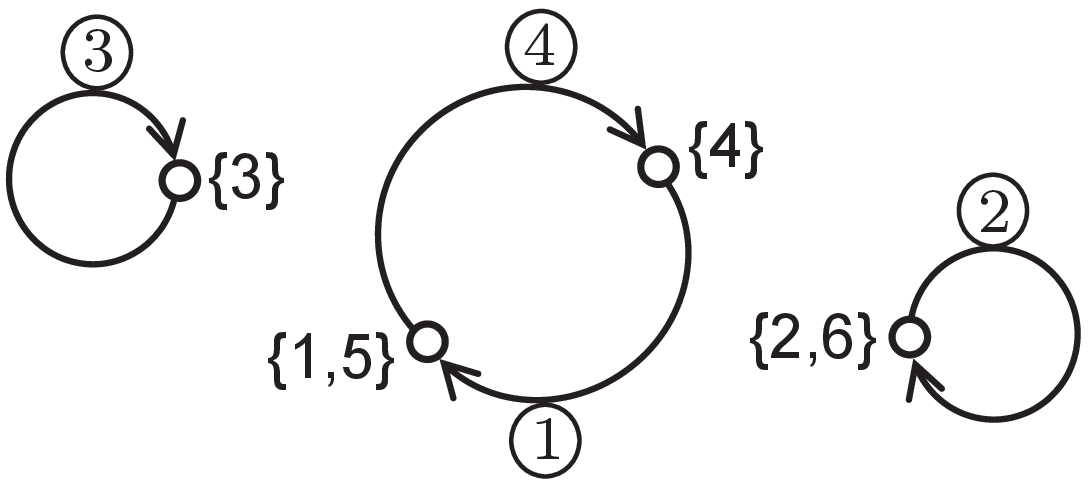}
		\caption{The quotient graph of the graph in Figure~\ref{subfig:odd-rep}.} 
		\label{subfig:odd-quotient}
	\end{subfigure}
	\hfil  \hfil  \hfil  \hfil  \hfil  \hfil  \hfil
	\begin{subfigure}{0.311\textwidth}
		\includegraphics[width=\linewidth]{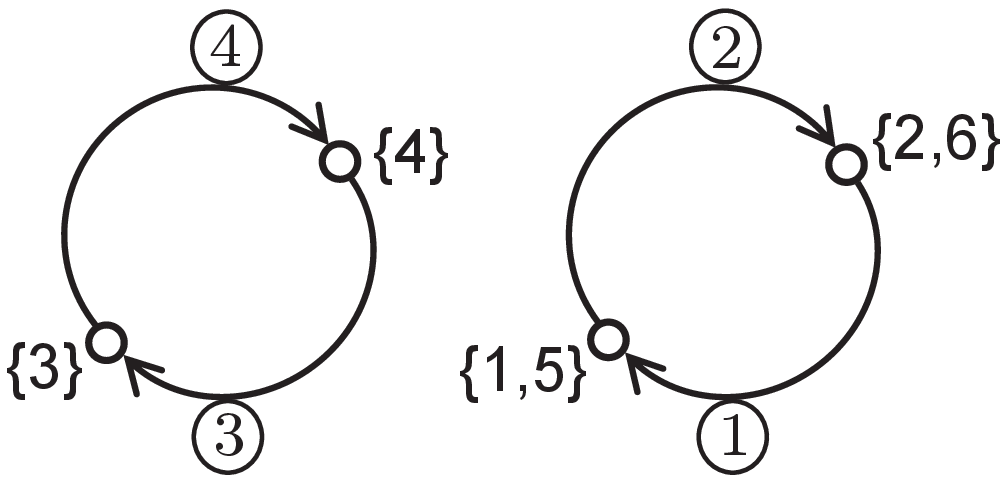}
		\caption{The quotient graph of the graph in Figure~\ref{subfig:even-rep}.} 
		\label{subfig:even-quotient}
	\end{subfigure}
	\caption{} 	
	\label{fig:odd-even-quotient}
\end{figure*}

Two multi-colored subgraphs $\mathbb{S}_1, \mathbb{S}_2 \in \mathcal{R}(\mathbb{G})$ are called \emph{similar} if $\mathbb{S}_1$ and $\mathbb{S}_2$ have the same $m$ sink vertices and the same set of $n$ colors. Graph similarity is an equivalence relation on $\mathcal{R}(\mathbb{G})$. The corresponding equivalence classes induced by this relation are called \emph{similarity classes}.

As an example of this concept, let $\mathbb{G}$ be the graph in Figure~\ref{fig:graph-example}. Let $\mathcal{E}_1$ be the similarity class of multi-colored subgraphs with sink vertices $1$ and $2$, and colors 1, 2, 3, 4. Figure~\ref{subfig:odd-rep} and Figure~\ref{subfig:even-rep} show the two multi-colored subgraphs in $\mathcal{E}_1$. Figure~\ref{subfig:single-rep} shows a multi-colored subgraph in the similarity class $\mathcal{E}_2$ with sink vertices $2$ and $6$, and colors 1, 3, 4, 5. In fact, this graph is the only multi-colored subgraph of $\mathbb{G}$ in $\mathcal{E}_2$.

Specific quotient graphs of the multi-colored subgraphs in the same similarity class will be used to define an important property of the class. Let $\mathcal{V}$ be the vertex set of a structural controllability graph $\mathbb{G}$. For any subset $\mathcal{U} \subset \mathcal{V}$, let $|\mathcal{U}|$ be the number of elements in $\mathcal{U}$. Let $\mathcal{V}_{\text{source}} \subset \mathcal{V}$ be the set of $m$ source vertices of every multi-colored subgraph of $\mathbb{G}$. Let $\mathcal{V}_{\text{sink}} \subset \mathcal{V}$ be the set of $m$ sink vertices of a given multi-colored subgraph $\mathbb{S}$ of $\mathbb{G}$. So $\mathcal{V}_{\text{source}} \cap \mathcal{V}_{\text{sink}}$ is the set of isolated vertices in $\mathbb{S}$ and $|\mathcal{V}_{\text{source}} - (\mathcal{V}_{\text{source}} \cap \mathcal{V}_{\text{sink}})| = |\mathcal{V}_{\text{sink}} - (\mathcal{V}_{\text{source}} \cap \mathcal{V}_{\text{sink}})| \leq m$. The desired quotient graph of $\mathbb{S}$ is induced by a ``matrimonial partition". A partition $\pi$ of $\mathcal{V}$ is a \emph{matrimonial partition} for $\mathbb{S}$ if it pairs each vertex in $\mathcal{V}_{\text{source}} - (\mathcal{V}_{\text{source}} \cap \mathcal{V}_{\text{sink}})$ with a different vertex in $\mathcal{V}_{\text{sink}} - (\mathcal{V}_{\text{source}} \cap \mathcal{V}_{\text{sink}})$ and assigns each pair to a different cell, then assigns each of the rest vertices in $\mathcal{V}$ to a new cell. So there are $|\mathcal{V}| - |\mathcal{V}_{\text{source}} - (\mathcal{V}_{\text{source}} \cap \mathcal{V}_{\text{sink}})|$ cells in $\pi$ and each of them has at most two vertices. If the pairing is not unique, $\pi$ is not unique.

An observation made by comparing $\mathbb{S}$ and the quotient graph $\mathbb{S}/\pi$ is that the cycle graphs and isolated vertices in $\mathbb{S}$ remain the same in $\mathbb{S}/\pi$, while the path graphs with positive lengths in $\mathbb{S}$ are, roughly speaking, ``welded" together to form new cycle graphs in $\mathbb{S}/\pi$. In the sequel, it is assumed that the quotient graphs of all multi-colored subgraphs in one similarity class are induced by the same matrimonial partition.

For example, $\pi = \{\{1, 5\}, \{2, 6\}, \{3\}, \{4\}\}$ is a matrimonial partition for the two multi-colored subgraphs in Figure~\ref{subfig:odd-rep} and Figure~\ref{subfig:even-rep}. The quotient graphs of the two graphs induced by $\pi$ are shown in Figure~\ref{subfig:odd-quotient} and Figure~\ref{subfig:even-quotient}, respectively.

A multi-colored subgraph is \emph{odd} (respectively, \emph{even}) if its quotient graph induced by a matrimonial partition has an odd (respectively, even) number of cycle graphs. As will be stated in Lemma~\ref{lem:bijctn-parity}, the choice of the matrimonial partition does not affect the relative parity of two multi-colored subgraphs in the same similarity class as long as their quotient graphs are induced by the same partition, where parity is the property of being odd or even. A similarity class of multi-colored subgraphs is \emph{balanced} if the numbers of odd and even multi-colored subgraphs in the similarity class are equal. Otherwise, it is \emph{unbalanced}. This important property of a similarity class is regardless of which matrimonial partition is chosen.

From Figure~\ref{fig:odd-even-quotient}, one knows that the graph in Figure~\ref{subfig:odd-rep} is odd and the graph in Figure~\ref{subfig:even-rep} is even, so the similarity class $\mathcal{E}_1$ is balanced. The similarity class $\mathcal{E}_2$ is unbalanced as it only has one multi-colored subgraph.

A ``cactus graph" introduced by Lin is a weakly connected graph consisting of one ``trunk" and any number of ``buds". A \emph{trunk} is a path graph with at least one vertex. A \emph{bud} consists of one cycle graph with at least one vertex and one additional arc called the bud's \emph{stem} which is incident to one of the cycle's vertices. A \emph{cactus graph} is then a weakly connected graph with exactly one trunk and any non-negative number of buds with the understanding that the stem of each bud comes out of either a vertex on the trunk or a vertex on the cycle of another bud in the graph. In this context a path graph is a cactus graph with no bud. Some definition of a cactus graph requires that the stem of a bud cannot come out of the last vertex on the trunk, but the definition in this paper does not, because it does not matter. The graphical condition involving cactus graphs is always that the original graph has a spanning subgraph which is a cactus graph or a disjoint union of cactus graphs. If the original graph has a spanning subgraph that is a cactus graph with one bud whose stem comes out of the last vertex on the trunk, the original graph must also have a spanning subgraph that is a cactus graph with no bud, obtained by removing a specific arc in the cycle of the bud, which points toward the same vertex as the stem of the bud does. So both definitions work. Note that a cactus graph has a unique root vertex. The condensation of a cactus graph which results when all cycles are condensed into vertices is a directed rooted tree.

\begin{figure}[!h]
    \centering
    \includegraphics[width=0.48\textwidth]{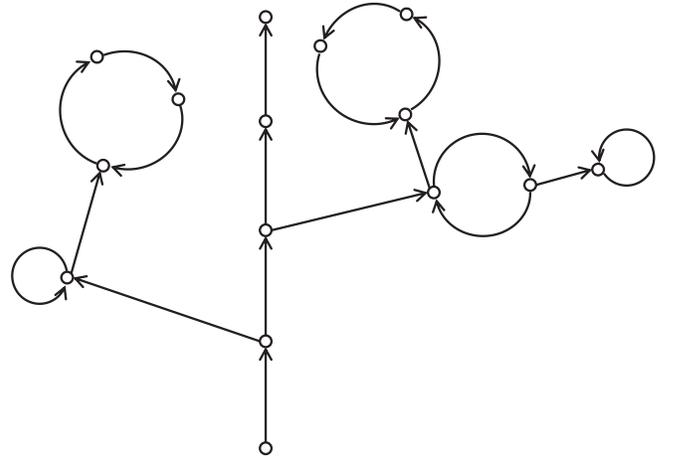}
    \caption{A cactus graph.}
    \label{fig:cactus-graph}
\end{figure}

Figure~\ref{fig:cactus-graph} gives an example of a cactus graph with five buds.

\subsection{Algebraic Concepts}

The \emph{generic rank} of a linearly parameterized matrix 
\begin{equation} \label{eqn:lnr-prm-mtrx}
M(p) = \sum_{k \in \mathbf{q}} g_k p_k h_k
\end{equation}
denoted by $\g-rank M$, is the maximum rank of $M$ that can be achieved as $p$ varies over ${\rm I\!R}^q$. It is generic in the sense that it is achievable by any $p$ in the complement of a proper algebraic set in ${\rm I\!R}^q$. Generalizing the standard notion of irreducibility, a matrix pair $(A, B)$ is said to be \emph{irreducible} if there is no permutation matrix $\Pi$ bringing $(A, B)$ into the form
\begin{equation*}
\Pi A \Pi^{-1} =
\begin{bmatrix}
A_1 & \mathbf{0} \\
A_2 & A_3
\end{bmatrix}, \quad
\Pi B =
\begin{bmatrix}
\mathbf{0} \\
B_1
\end{bmatrix}
\end{equation*}
where $A_1$ is an $n_1 \times n_1$ block, $B_1$ is an $(n-n_1) \times m$ block, $1 \leq n_1 < n$.

\begin{proposition} \textnormal{\cite{mayeda1981structural}}  \label{prp:irrdc-forest}
A linearly parameterized matrix pair $(A, B)$ is irreducible if and only if the graph of $(A, B)$ has a spanning forest rooted at the $m$ vertices with labels $n+1$ to $n+m$.
\end{proposition}

Although Proposition~\ref{prp:irrdc-forest} was initially developed for matrix pairs satisfying the unitary assumption, the same proof applies to all linearly parameterized matrix pairs without change. Therefore a proof of Proposition~\ref{prp:irrdc-forest} will not be given here.

\section{Main Result}

The following classical result characterizes the structural controllability of linearly parameterized matrix pairs satisfying the unitary assumption.

\begin{proposition} \textnormal{\cite{lin1974structural, shields1975structural, mayeda1981structural}}  \label{prp:uni-prm-cond}
Let $(A, B)$ be a linearly parameterized matrix pair which satisfies the unitary assumption. The following statements are equivalent. \\
(\lowercase\expandafter{\romannumeral1}) The pair $(A, B)$ is structurally controllable. \\
(\lowercase\expandafter{\romannumeral2}) $\g-rank [A \enspace B] = n$ and $(A, B)$ is irreducible. \\
(\lowercase\expandafter{\romannumeral3}) The graph of $(A, B)$ has a spanning subgraph which is a disjoint union of $m$ cactus graphs rooted at the $m$ vertices with labels $n+1$ to $n+m$, respectively.
\end{proposition}

The graphical conditions in Proposition~\ref{prp:uni-prm-cond} is equivalent to the graphical conditions given in \cite{liu2011controllability} for structural controllability. A ``maximum matching" defined in \cite{liu2011controllability} is a maximum-cardinality set of arcs that do not share start vertices or end vertices. It will be called a \emph{nonstandard maximum matching} in the rest of the paper because it differs from the standard definition of maximum matching, i.e., a maximum-cardinality set of arcs that do not share vertices, in the sense that a nonstandard matching allows the start vertex of an arc to be the end vertex of another arc, but a standard matching does not. Let $(A, B)$ be a linearly parameterized matrix pair which satisfies the unitary assumption. The following statements are equivalent. \\
(\lowercase\expandafter{\romannumeral1}) $\g-rank [A \enspace B] = n$. \\
(\lowercase\expandafter{\romannumeral2}) The graph of $(A, B)$ has a spanning subgraph which is a disjoint union of $m$ path graphs and any number of cycle graphs. \\
(\lowercase\expandafter{\romannumeral3}) The graph of $(A, B)$ has a nonstandard maximum matching of size $n$. \\
The equivalence of (\lowercase\expandafter{\romannumeral1}) and (\lowercase\expandafter{\romannumeral2}) is given by Lemma~2 in \cite{mayeda1981structural}. The equivalence of (\lowercase\expandafter{\romannumeral1}) and (\lowercase\expandafter{\romannumeral3}) is established as follows. It is possible to represent the graph $\mathbb{G}$ of $(A, B)$ by a bipartite graph $\mathbb{B}$ such that each vertex $i$ of $\mathbb{G}$ becomes two vertices $i^+$ and $i^-$ in $\mathbb{B}$ and each arc $(j, i)$ of $\mathbb{G}$ corresponds to an arc $(j^+, i^-)$ in $\mathbb{B}$. Lemma~1 in \cite{commault2002characterization} implies that $\g-rank [A \enspace B] = n$ if and only if $\mathbb{B}$ has a standard maximum matching of size $n$. It is easy to see that a standard maximum matching in $\mathbb{B}$ corresponds to a nonstandard maximum matching in $\mathbb{G}$. So (\lowercase\expandafter{\romannumeral1}) and (\lowercase\expandafter{\romannumeral3}) are equivalent. Between the two graphical conditions for generic rank, (\lowercase\expandafter{\romannumeral2}) is easier to visualize in $\mathbb{G}$ and to combine with the graphical condition for irreducibility.

The following theorem, which is the main result of this paper, shows how the graphical condition in Proposition~\ref{prp:uni-prm-cond} changes when the unitary assumption is relaxed to the binary assumption.

\begin{theorem}  \label{thm:bin-prm-cond}
Let $(A, B)$ be a linearly parameterized matrix pair which satisfies the binary assumption. The following statements are equivalent. \\
(\lowercase\expandafter{\romannumeral1}) The pair $(A, B)$ is structurally controllable. \\
(\lowercase\expandafter{\romannumeral2}) $\g-rank [A \enspace B] = n$ and $(A, B)$ is irreducible. \\
(\lowercase\expandafter{\romannumeral3}) The graph of $(A, B)$ has an unbalanced similarity class of multi-colored subgraphs and has a spanning subgraph which is a disjoint union of $m$ cactus graphs rooted at the $m$ vertices with labels $n+1$ to $n+m$, respectively. \\
(\lowercase\expandafter{\romannumeral4}) The graph of $(A, B)$ has an unbalanced similarity class of multi-colored subgraphs and has a spanning forest rooted at the $m$ vertices with labels $n+1$ to $n+m$.
\end{theorem}

When subject to the unitary assumption, Theorem~\ref{thm:bin-prm-cond} reduces to Proposition~\ref{prp:uni-prm-cond}. To understand why this is so, let $\mathbb{G}$ be the graph of a matrix pair $(A, B)$ which satisfies the unitary assumption. As no two arcs of $\mathbb{G}$ are of the same color, $\mathbb{G}$ has an unbalanced similarity class of multi-colored subgraphs if and only if $\mathbb{G}$ has a multi-colored subgraph, which can be obtained by removing the stems of all buds in the $m$ cactus graphs. So condition (\lowercase\expandafter{\romannumeral3}) in Theorem~\ref{thm:bin-prm-cond} reduces to condition (\lowercase\expandafter{\romannumeral3}) in Proposition~\ref{prp:uni-prm-cond}.

As an example of Theorem~\ref{thm:bin-prm-cond}, the matrix pair given in (\ref{eqn:bin-prm-exmp}) is structurally controllable because the graph in Figure~\ref{fig:graph-example} satisfies condition (\lowercase\expandafter{\romannumeral4}).

\section{Analysis}

This section focuses on the analysis and proof of Theorem~\ref{thm:bin-prm-cond}, in which the equivalence of statements (\lowercase\expandafter{\romannumeral1}) and (\lowercase\expandafter{\romannumeral2}) is proved first, followed by the equivalence of statements (\lowercase\expandafter{\romannumeral2}) and (\lowercase\expandafter{\romannumeral4}), and then that of statements (\lowercase\expandafter{\romannumeral3}) and (\lowercase\expandafter{\romannumeral4}).

\subsection{Proof of Theorem~\ref{thm:bin-prm-cond}, (\lowercase\expandafter{\romannumeral1})$\iff$(\lowercase\expandafter{\romannumeral2})}

Apparently, if a linearly parameterized matrix pair $(A, B)$ is structurally controllable, $(A, B)$ is irreducible and $\g-rank [A \enspace B] = n$. We will prove the converse. To do that, some concepts and certain result from \cite{corfmat1976structurally} are summarized as they apply to the proof. It is worth pointing out that the concepts and the result in \cite{corfmat1976structurally} do not require the binary assumption.

Suppose $\mathcal{S} = \{i_1, i_2, \dots, i_k\} \subset \mathbf{q}$ with $i_1 < i_2 < \dots < i_k$. Let matrices $G_{\mathcal{S}}$, $H_{\mathcal{S}}$ and $P_{\mathcal{S}}$ be
\begin{align*}
G_{\mathcal{S}} & \triangleq
\begin{bmatrix}
g_{i_1} & g_{i_2} & \dots & g_{i_k}
\end{bmatrix}, \quad
H_{\mathcal{S}} \triangleq
\begin{bmatrix}
h_{i_1} \\
h_{i_2} \\
\vdots \\
h_{i_k}
\end{bmatrix} \\
P_{\mathcal{S}} & \triangleq \diag \{p_{i_1}, p_{i_2}, \dots, p_{i_k}\}
\end{align*}
If $\mathcal{S} = \emptyset$, $G_{\mathcal{S}}$, $H_{\mathcal{S}}$ and $P_{\mathcal{S}}$ are each the $0 \times 0$ matrix. The complement of $\mathcal{S}$ in $\mathbf{q}$ is denoted by $\mathbf{q} - \mathcal{S}$. Note that the linear parameterization $\sum_{i \in \mathbf{q}} g_i p_i h_i$ is exactly $G_{\mathbf{q}} P_{\mathbf{q}} H_{\mathbf{q}}$.

The \emph{transfer matrix} of $\{G_{\mathbf{q}}, H_{\mathbf{q}}\}$, denoted by $T$, is a block matrix with $q$ row partitions and $q+1$ column partitions defined as
\begin{equation*}
T_{i,j} = 
\left\lbrace
\begin{array}{ll}
h_{i1} g_j ,     &  i,j \in \mathbf{q} \\
h_{i2} , &  i \in \mathbf{q}, j=0 
\end{array}
\right.
\end{equation*}
where $T_{i,j}$ is the $ij$th block of $T$, $g_j \in {\rm I\!R}^n$, $h_{i1} \in {\rm I\!R}^{1 \times n}$ and $h_{i2} \in {\rm I\!R}^{1 \times m}$. The \emph{transfer graph} of $\{G_{\mathbf{q}}, H_{\mathbf{q}}\}$, written $\mathbb{T}$, is the graph of the transfer matrix $T$ and is defined to be an unweighted directed graph with $q+1$ vertices labeled $0$, $1$, $\dots$, $q$ and an arc from vertex $j$ to vertex $i$ whenever $T_{i,j}$ is nonzero. The following proposition is derived from Theorem~1 in \cite{corfmat1976structurally} with constant matrices $A_0 = \mathbf{0}$ and $B_0 = \mathbf{0}$. It is applicable to any linearly parameterized matrix pair with or without the binary assumption.

\begin{proposition} \textnormal{\cite{corfmat1976structurally}} \label{prp:lnr-prm-cond}
A linearly parameterized matrix pair $(A, B)$ given by (\ref{eqn:lnr-prm-pair-cmb}) is structurally controllable if and only if 
\begin{equation} \label{eqn:min-rank}
\min\limits_{\mathcal{S} \subset \mathbf{q}} \, (\rank G_{\mathcal{S}} + \rank H_{\mathbf{q}-\mathcal{S}}) = n
\end{equation}
and the transfer graph of $\{G_{\mathbf{q}}, H_{\mathbf{q}}\}$ has a spanning tree rooted at vertex $0$.
\end{proposition}

In addition to Proposition~\ref{prp:lnr-prm-cond}, three lemmas are needed to prove the equivalence of statements (\lowercase\expandafter{\romannumeral1}) and (\lowercase\expandafter{\romannumeral2}) in Theorem~\ref{thm:bin-prm-cond}. More specifically, Lemma~\ref{lem:genrnk-min} and Lemma~\ref{lem:G-T} draw a connection between Proposition~\ref{prp:lnr-prm-cond} and statement (\lowercase\expandafter{\romannumeral2}). The following concepts and Lemma~\ref{lem:max-min} are the key ideas for proving Lemma~\ref{lem:genrnk-min}. Among the three lemmas, Lemma~\ref{lem:max-min} and Lemma~\ref{lem:genrnk-min} hold without the binary assumption, but Lemma~\ref{lem:G-T} needs the binary assumption.

Suppose we are given two real matrices
\begin{equation*}
G_{n_1 \times k} =
\begin{bmatrix}
g_1 & g_2 & \dots & g_k
\end{bmatrix}, \quad
H_{k \times n_2} =
\begin{bmatrix}
h_1 \\
h_2 \\
\vdots \\
h_k
\end{bmatrix}
\end{equation*}
Let $\mathbf{k} \triangleq \{1, 2, \dots, k\}$. Let $\mathcal{P} \triangleq \{(g_i, h_i) \, | \, i \in \mathbf{k}\}$ be the set of $k$ pairs of vectors. For $\mathcal{I} \subset \mathbf{k}$, a nonempty subset $\{(g_i, h_i) \, | \, i \in \mathcal{I}\} \subset \mathcal{P}$ is \emph{jointly independent} if $\{g_i \, | \, i \in \mathcal{I}\}$ and $\{h_i \, | \, i \in \mathcal{I}\}$ are both linearly independent sets. That is, 
\begin{equation*}
\rank G_{\mathcal{I}} = \rank H_{\mathcal{I}} = |\mathcal{I}|
\end{equation*}
where $|\mathcal{I}|$ is the cardinality of $\mathcal{I}$, i.e., the number of elements in $\mathcal{I}$. Then $\mathcal{I}$ is called a \emph{jointly independent index set} of $(G, H)$. Let $\mathcal{J}(G, H)$ be the set of all jointly independent index sets of $(G, H)$.

\begin{lemma} \label{lem:max-min}
For a linearly parameterized matrix $M$ given by (\ref{eqn:lnr-prm-mtrx}), 
\begin{equation*}
\max\limits_{\mathcal{I} \in \mathcal{J}(G_{\mathbf{q}}, H_{\mathbf{q}})} |\mathcal{I}| = 
\min\limits_{\mathcal{S} \subset \mathbf{q}} \, (\rank G_{\mathcal{S}} + \rank H_{\mathbf{q}-\mathcal{S}})
\end{equation*}
\end{lemma}

\noindent \textbf{Proof of Lemma~\ref{lem:max-min}:} Let $\mathcal{M}_1 = \{\mathbf{q}, \mathcal{J}_1\}$ and $\mathcal{M}_2 = \{\mathbf{q}, \mathcal{J}_2\}$ be two finite matroids, where $\mathbf{q}$ is the ground set; $\mathcal{J}_1$ is the family of the independent sets of $\mathbf{q}$ defined by the linear independence relation of $\{g_i \, | \, i \in \mathbf{q}\}$, i.e., $\mathcal{S}_1 \in \mathcal{J}_1$ if and only if $\{g_i \, | \, i \in \mathcal{S}_1\}$ is a linearly independent set; $\mathcal{J}_2$ is the family of the independent sets of $\mathbf{q}$ defined by the linear independence relation of $\{h_i \, | \, i \in \mathbf{q}\}$, i.e., $\mathcal{S}_2 \in \mathcal{J}_2$ if and only if $\{h_i \, | \, i \in \mathcal{S}_2\}$ is a linearly independent set. Let $r_1$ and $r_2$ be the rank functions of $\mathcal{M}_1$ and $\mathcal{M}_2$, respectively. Naturally, $\forall \mathcal{S}_1 \subset \mathbf{q}$, $r_1(\mathcal{S}_1) = \rank G_{\mathcal{S}_1}$, and $\forall \mathcal{S}_2 \subset \mathbf{q}$, $r_2(\mathcal{S}_2) = \rank H_{\mathcal{S}_2}$. By the matroid intersection theorem \cite{murota2010matroids}, 
\begin{equation*}
\max\limits_{\mathcal{I} \in \mathcal{J}_1 \cap \mathcal{J}_2} |\mathcal{I}| = 
\min\limits_{\mathcal{S} \subset \mathbf{q}} \left( r_1(\mathcal{S}) + r_2(\mathbf{q}-\mathcal{S}) \right)
\end{equation*}
That is,
\begin{equation*}
\max\limits_{\mathcal{I} \in \mathcal{J}(G_{\mathbf{q}}, H_{\mathbf{q}})} |\mathcal{I}| = 
\min\limits_{\mathcal{S} \subset \mathbf{q}} \, (\rank G_{\mathcal{S}} + \rank H_{\mathbf{q}-\mathcal{S}}) 
\end{equation*}
Therefore, Lemma~\ref{lem:max-min} is true. \hfill $\qed$

\begin{lemma} \label{lem:genrnk-min}
For a linearly parameterized matrix $M$ given by (\ref{eqn:lnr-prm-mtrx}),
\begin{equation*}
\g-rank M = \min\limits_{\mathcal{S} \subset \mathbf{q}} \, (\rank G_{\mathcal{S}} + \rank H_{\mathbf{q}-\mathcal{S}})
\end{equation*}
\end{lemma}

\noindent \textbf{Proof of Lemma~\ref{lem:genrnk-min}:} Let $\mathcal{I}$ be a jointly independent index set of $\{G_{\mathbf{q}}, H_{\mathbf{q}}\}$ with the maximum cardinality. Let $p_i = 1$ if $i \in \mathcal{I}$ and $p_i = 0$ if $i \notin \mathcal{I}$. Then $M = G_{\mathcal{I}} H_{\mathcal{I}}$. As $\rank G_{\mathcal{I}} = \rank H_{\mathcal{I}} = |\mathcal{I}|$, $\rank G_{\mathcal{I}} H_{\mathcal{I}} = |\mathcal{I}|$. So 
\begin{equation*}
\g-rank M \geq \rank G_{\mathcal{I}} H_{\mathcal{I}} = |\mathcal{I}|
\end{equation*} 
For any $\mathcal{S} \subset \mathbf{q}$, $G_{\mathbf{q}} P_{\mathbf{q}} H_{\mathbf{q}} = G_{\mathcal{S}} P_{\mathcal{S}} H_{\mathcal{S}} + G_{\mathbf{q}-\mathcal{S}} P_{\mathbf{q}-\mathcal{S}} H_{\mathbf{q}-\mathcal{S}}$. So
\begin{align*}
\rank G_{\mathbf{q}} P_{\mathbf{q}} H_{\mathbf{q}}
& \leq \rank G_{\mathcal{S}} P_{\mathcal{S}} H_{\mathcal{S}} + \rank G_{\mathbf{q}-\mathcal{S}} P_{\mathbf{q}-\mathcal{S}} H_{\mathbf{q}-\mathcal{S}} \\
& \leq \rank G_{\mathcal{S}} + \rank H_{\mathbf{q}-\mathcal{S}}
\end{align*}
holds for all $p \in {\rm I\!R}^q$, $\mathcal{S} \subset \mathbf{q}$. It follows by varying $p$ over ${\rm I\!R}^q$ on the left side of the inequality and by varying $\mathcal{S}$ over the power set of $\mathbf{q}$ on the right side of the inequality that
\begin{equation*}
\g-rank M \leq \min\limits_{\mathcal{S} \subset \mathbf{q}} \, (\rank G_{\mathcal{S}} + \rank H_{\mathbf{q}-\mathcal{S}})
\end{equation*}
By Lemma~\ref{lem:max-min}, $|\mathcal{I}| = \min\limits_{\mathcal{S} \subset \mathbf{q}} \, (\rank G_{\mathcal{S}} + \rank H_{\mathbf{q}-\mathcal{S}})$. So 
\begin{equation*}
\g-rank M = |\mathcal{I}| = \min\limits_{\mathcal{S} \subset \mathbf{q}} \, (\rank G_{\mathcal{S}} + \rank H_{\mathbf{q}-\mathcal{S}}) 
\end{equation*}
Therefore, Lemma~\ref{lem:genrnk-min} is true. \hfill $\qed$

\begin{corollary} \label{crl:rankAB}
For real matrices $G_{n_1 \times k}$ and $H_{k \times n_2}$, 
\begin{equation*}
\rank GH \leq \max\limits_{\mathcal{I} \in \mathcal{J}(G, H)} |\mathcal{I}|
\end{equation*}
\end{corollary}

\noindent \textbf{Proof of Corallary~\ref{crl:rankAB}:} By Lemma~\ref{lem:max-min} and Lemma~\ref{lem:genrnk-min}, 
\begin{equation*}
\g-rank G P_{\mathbf{k}} H = \max\limits_{\mathcal{I} \in \mathcal{J}(G, H)} |\mathcal{I}|
\end{equation*}
So
\begin{align*}
\rank G H = \rank G I H 
&\leq \g-rank G P_{\mathbf{k}} H \\
&= \max\limits_{\mathcal{I} \in \mathcal{J}(G, H)} |\mathcal{I}| 
\end{align*}
Therefore, Corallary~\ref{crl:rankAB} is true. \hfill $\qed$

Corollary~\ref{crl:rankAB} gives a tighter upper bound on $\rank G H$ than $\min \, \{\rank G, \rank H\}$.

The concept of ``line graph" is useful for proving Lemma~\ref{lem:G-T}. The \emph{line graph} of a given structural controllability graph $\mathbb{G}$, written $\mathbb{L}(\mathbb{G})$, is an unweighted directed graph that has one vertex for each arc of $\mathbb{G}$, for example a vertex $ijk$ for an arc $(i, j)_k$ in $\mathbb{G}$, and has an arc from vertex $abk_1$ to vertex $bck_2$ if $\mathbb{G}$ has arcs $(a, b)_{k_1}$ and $(b, c)_{k_2}$. That is, each arc in $\mathbb{L}(\mathbb{G})$ represents a length-two walk \cite{godsil2013algebraic} in $\mathbb{G}$.

\begin{figure}[!h]
	\centering
	\begin{subfigure}{0.076\textwidth}
		\includegraphics[width=\linewidth]{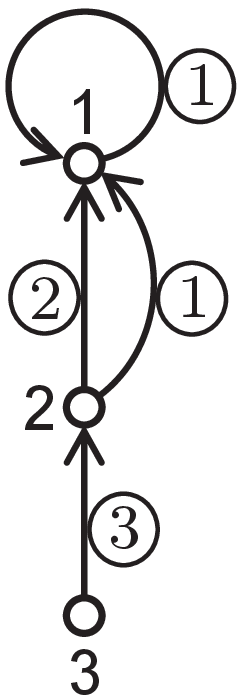}
		\caption{$\mathbb{G}$} 
		\label{subfig:graph}
	\end{subfigure}
	\hfil  \hfil  \hfil  \hfil  \hfil
	\begin{subfigure}{0.148\textwidth}
		\includegraphics[width=\linewidth]{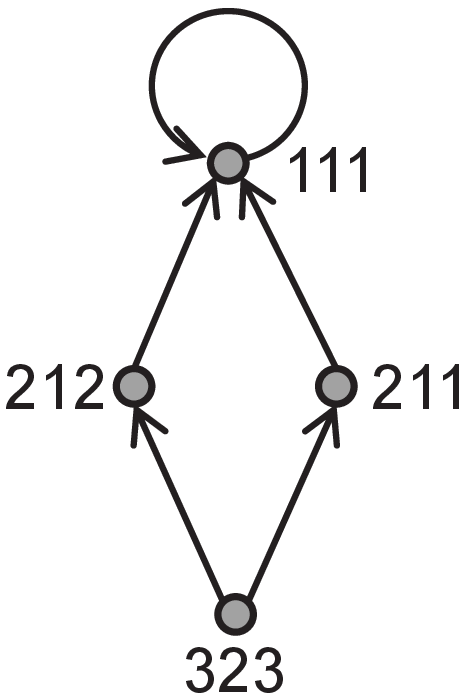}
		\caption{$\mathbb{L}(\mathbb{G})$} 
		\label{subfig:line-graph}
	\end{subfigure}
	\caption{}    
	\label{fig:line-graph-exmple}
\end{figure}

Figure~\ref{fig:line-graph-exmple} gives an example of a structural controllability graph and its line graph.

\begin{lemma} \label{lem:G-T}
Let $(A, B)$ be a linearly parameterized matrix pair given by (\ref{eqn:lnr-prm-pair-cmb}), which satisfies the binary assumption. If $(A, B)$ is irreducible, the transfer graph of $\{G_{\mathbf{q}}, H_{\mathbf{q}}\}$ has a spanning tree rooted at vertex $0$.
\end{lemma}

\noindent \textbf{Proof of Lemma~\ref{lem:G-T}:} For clarity, let $v_i$ denote vertex $i$ in the graph $\mathbb{G}$ of $(A, B)$ and let $w_i$ denote vertex $i$ in the transfer graph $\mathbb{T}$ of $\{G_{\mathbf{q}}, H_{\mathbf{q}}\}$. Let $\mathcal{B} \triangleq \{i \in \mathbf{q} \, | \, h_{i2} \neq \mathbf{0}\}$. In other words, $p_i$ appears in $B$ if and only if $i \in \mathcal{B}$. By definition, there is an arc in $\mathbb{T}$ from vertex $w_0$ to vertex $w_i$ for each $i \in \mathcal{B}$. For $i, j \in \mathbf{q}$, there is an arc in $\mathbb{T}$ from $w_i$ to $w_j$ if $T_{j,i} = h_{j1} g_i \neq 0$. As $h_{j1}$ and $g_i$ are binary vectors, $h_{j1} g_i \neq 0$ if and only if $\exists \, k \in \{1, 2, \dots, n\}$ such that the $k$th entry of $h_{j1}$ is one and the $k$th entry of $g_i$ is also one. So $\mathbb{T}$ has an arc from $w_i$ to $w_j$ if and only if $\mathbb{G}$ has an arc of color $i$ pointing toward $v_k$ and an arc of color $j$ leaving $v_k$.

Let $\widehat{\mathbb{T}}$ be the subgraph of $\mathbb{T}$ induced by vertices $w_1$, $w_2$, $\dots$, $w_q$. Remember that the line graph $\mathbb{L}(\mathbb{G})$ has one vertex for each arc of $\mathbb{G}$. Let $\pi$ be the partition of the vertices of $\mathbb{L}(\mathbb{G})$ such that the vertices for the arcs of $\mathbb{G}$ in the same color are in the same cell of the partition. Obviously, the quotient graph $\mathbb{L}(\mathbb{G})/\pi$ has $q$ vertices. For each $i \in \mathbf{q}$, let $u_i$ denote vertex $i$ in $\mathbb{L}(\mathbb{G})/\pi$, which corresponds to the arcs of $\mathbb{G}$ with color $i$. Then $\mathbb{L}(\mathbb{G})/\pi$ and $\widehat{\mathbb{T}}$ are isomorphic with the bijection that maps vertex $u_i$ in $\mathbb{L}(\mathbb{G})/\pi$ to vertex $w_i$ in $\widehat{\mathbb{T}}$.

If $(A, B)$ is irreducible, by Proposition~\ref{prp:irrdc-forest}, $\mathbb{G}$ has a spanning forest rooted at the $m$ vertices $v_{n+1}$, $v_{n+2}$, $\dots$, and $v_{n+m}$. So $\mathbb{L}(\mathbb{G})$ has a spanning forest rooted at the vertices for the arcs of $\mathbb{G}$ leaving $v_{n+1}$, $v_{n+2}$, $\dots$, or $v_{n+m}$. The isomorphism of $\mathbb{L}(\mathbb{G})/\pi$ and $\widehat{\mathbb{T}}$ implies that $\widehat{\mathbb{T}}$ has a spanning forest rooted at the vertices in the set $\{w_i \, | \, i \in \mathcal{B}\}$. Since the transfer graph $\mathbb{T}$ has an arc from $w_0$ to $w_i$ for each $i \in \mathcal{B}$, $\mathbb{T}$ has a spanning tree rooted at $w_0$. \hfill $\qed$

\noindent \textbf{Proof of Theorem~\ref{thm:bin-prm-cond}, (\lowercase\expandafter{\romannumeral1})$\iff$(\lowercase\expandafter{\romannumeral2}):} Apparently, (\lowercase\expandafter{\romannumeral1}) $\Longrightarrow$ (\lowercase\expandafter{\romannumeral2}). If (\lowercase\expandafter{\romannumeral2}) is true, by Lemma~\ref{lem:genrnk-min} and Lemma~\ref{lem:G-T}, 
\begin{equation*}
\min\limits_{\mathcal{S} \subset \mathbf{q}} \, (\rank G_{\mathcal{S}} + \rank H_{\mathbf{q}-\mathcal{S}}) = n 
\end{equation*}
and the transfer graph of $\{G_{\mathbf{q}}, H_{\mathbf{q}}\}$ has a spanning tree rooted at vertex $0$. By Proposition~\ref{prp:lnr-prm-cond}, (\lowercase\expandafter{\romannumeral1}) is true. So (\lowercase\expandafter{\romannumeral1})$\iff$(\lowercase\expandafter{\romannumeral2}). \hfill $\qed$

\subsection{Proof of Theorem~\ref{thm:bin-prm-cond}, (\lowercase\expandafter{\romannumeral2})$\iff$(\lowercase\expandafter{\romannumeral4})}

Two lemmas are needed to prove the equivalence of statements (\lowercase\expandafter{\romannumeral2}) and (\lowercase\expandafter{\romannumeral4}). Lemma~\ref{lem:bijctn-parity} implies that the balance or unbalance of a similarity class of multi-colored subgraphs is an intrinsic property, regardless of which matrimonial partition is chosen. It facilitates the understanding of Lemma~\ref{lem:genrnk-graph}, which converts the generic rank condition into a graphical condition.

For the proofs of Lemma~\ref{lem:bijctn-parity} and Lemma~\ref{lem:genrnk-graph}, some bases of permutation and determinant are needed. Let $\mathcal{S}_n$ be the set of all permutations of the set $\mathbf{n} \triangleq \{1, 2, \dots, n\}$. Let $\sigma \in \mathcal{S}_n$ be one such permutation which maps $i \in \mathbf{n}$ to $\sigma(i) \in \mathbf{n}$. $\sigma$ is \emph{odd} (respectively, \emph{even}) if $\sigma(1)$, $\sigma(2)$, $\dots$, $\sigma(n)$ can be transformed into 1, 2, $\dots$, $n$ by an odd (respectively, even) number of two-element swaps. The \emph{signature} of $\sigma$, denoted by $\sgn(\sigma)$, takes value from $\{1, -1\}$ such that $\sgn(\sigma) = 1$ if $\sigma$ is even, and $\sgn(\sigma) = -1$ if $\sigma$ is odd. Each permutation in $\mathcal{S}_n$ can be decomposed into the product of disjoint cycles. Let $c$ be the number of disjoint cycles that $\sigma$ can be decomposed into, then $\sigma$ is odd (respectively, even) if $n-c$ is odd (respectively, even). The composition of two permutations with the same parity (respectively, opposite parities) is an even (respectively, odd) permutation.

One definition of the determinant of an $n \times n$ square matrix $M$ is
\begin{equation} \label{eqn:det-def}
\det M = \sum_{\sigma \in \mathcal{S}_n} \sgn(\sigma) \prod_{i \in \mathbf{n}} m_{i, \sigma(i)}
\end{equation}
where $m_{i,j}$ is the $ij$th entry of $M$.

If $M_{n \times n}$ is linearly parameterized as given by (\ref{eqn:lnr-prm-mtrx}), the graph of $M$, written $\mathbb{G}_M$, is an unweighted directed graph with $n$ vertices labeled $1$ to $n$ and an arc $(j, i)_k$ if the $ij$th entry in $M$ contains $p_k$. So $\mathbb{G}_M$ is exactly the subgraph induced in the graph of the pair $(M, \mathbf{0})$ by the $n$ vertices with labels $1$ to $n$. With the binary assumption, each entry of $M$ is either zero, one parameter or the sum of distinct parameters. After the products of entries in (\ref{eqn:det-def}) are expanded, each term in $\det M$ is a signed product of $n$ parameters. As no two of the $n$ parameters in a term are taken from the same row or the same column of $M$, each term in $\det M$ corresponds to a spanning subgraph of $\mathbb{G}_M$ with $n$ arcs, which is a disjoint union of finite number of cycle graphs. The following proposition is derived from Theorem~2 in \cite{coates1959flow} and will be used to prove Lemma~\ref{lem:bijctn-parity}.

\begin{proposition} \textnormal{\cite{coates1959flow}}  \label{prp:sgn-in-det}
Let $M$ be an $n \times n$ linearly parameterized matrix satisfying the binary assumption, whose graph is denoted by $\mathbb{G}_M$. The sign of a term in $\det M$ is positive if $n-c$ is even, and is negative if $n-c$ is odd, where $c$ is the number of cycle graphs in the corresponding spanning subgraph of $\mathbb{G}_M$.
\end{proposition}

For an $n \times n$ linearly parameterized matrix $M$ which satisfies the binary assumption, a term in $\det M$ is \emph{valid} if it contains $n$ distinct parameters. Since $\det M$ is a multilinear function of $p_1$, $p_2$, $\dots$, $p_q$, only valid terms appear in the final expression of $\det M$. That is,
\begin{equation} \label{eqn:det-bin-prm}
\det M = \sum_{\substack{\mathcal{C} \subset \mathbf{q} \\ |\mathcal{C}|=n}} a_{\mathcal{C}} \prod_{k \in \mathcal{C}} p_k
\end{equation}
where $a_{\mathcal{C}} \in [-n! \, , n! \,]$ is the integer coefficient of the product of the $n$ distinct parameters labeled by elements of $\mathcal{C}$.

Let $\mathbb{G}$ be the graph of a linearly parameterized matrix pair $(A, B)$ which satisfies the binary assumption. By replacing any $m$ columns, such as columns $t_1$, $t_2$, $\dots$, $t_m$, of the partitioned matrix $[A \enspace B]$ with $\mathbf{0}$, we get another $n \times (n+m)$ matrix $[\widehat{A} \enspace \widehat{B}]$. The graph of the pair $(\widehat{A}, \widehat{B})$, denoted by $\widehat{\mathbb{G}}$, is then a spanning subgraph of $\mathbb{G}$ which results when all the arcs leaving vertices $t_1$, $t_2$, $\dots$, or $t_m$ are removed from $\mathbb{G}$. Let $M$ be the $n \times n$ submatrix obtained by deleting columns $t_1$, $t_2$, $\dots$, $t_m$ of $[A \enspace B]$. Each valid term in $\det M$ has $n$ distinct parameters and no two of them are taken from the same row or the same column of $[\widehat{A} \enspace \widehat{B}]$. So each valid term in $\det M$ corresponds to a spanning subgraph of $\widehat{\mathbb{G}}$ with $n$ arcs in $n$ distinct colors and with no two arcs pointing toward the same vertex or leaving the same vertex, which is a multi-colored subgraph of $\mathbb{G}$ with sink vertices $t_1$, $t_2$, $\dots$, $t_m$. Therefore, each valid term in the determinant of an $n \times n$ submatrix of $[A \enspace B]$ corresponds to a multi-colored subgraph of $\mathbb{G}$. Valid terms which are in the determinant of the same submatrix and which contain the same $n$ distinct parameters correspond to multi-colored subgraphs of $\mathbb{G}$ in the same similarity class.

\begin{lemma} \label{lem:bijctn-parity}
Let $\mathbb{G}$ be the graph of a linearly parameterized matrix pair $(A, B)$ which satisfies the binary assumption. The relative parity of two multi-colored subgraphs of $\mathbb{G}$ in the same similarity class remains unchanged regardless of which matrimonial partition induces their quotient graphs.
\end{lemma}

\noindent \textbf{Proof of Lemma~\ref{lem:bijctn-parity}:} Suppose $\mathbb{G}$ has a similarity class $\mathcal{E}$ of multi-colored subgraphs with sink vertices $t_1$, $t_2$, $\dots$, $t_m$. Let 
\begin{equation*}
\mathcal{L} \triangleq \{1, 2, \dots, n+m\} - \{t_1, t_2, \dots, t_m\}
\end{equation*} 
Note that $|\mathcal{L}| = n$. Let $M$ be the $n \times n$ submatrix obtained by deleting columns $t_1$, $t_2$, $\dots$, $t_m$ of $[A \enspace B]$. Let 
\begin{equation*}
f_0 \colon \mathbf{n} \to \mathcal{L} 
\end{equation*}
be the bijection such that for each $i \in \mathbf{n}$, the $i$th column of $M$ is taken from the $f_0(i)$th column of $[A \enspace B]$. Let $\pi$ be the matrimonial partition that induces the quotient graphs of all multi-colored subgraphs in $\mathcal{E}$. For each $i \in \mathcal{L} \cap \{n+1, n+2, \dots, n+m\}$, vertex $i$ is a source vertex but not a sink vertex. So vertex $i$ shares a cell of $\pi$ with a sink vertex, denoted by vertex $t_{\pi}^i$. Let
\begin{equation*}
f_{\pi}: \mathcal{L} \to \mathbf{n}
\end{equation*}
be the bijection such that $f_{\pi}(i) = t_{\pi}^i$ if $i \in \mathcal{L} \cap \{n+1, n+2, \dots, n+m\}$ and $f_{\pi}(i) = i$ if $i \in \mathcal{L} \cap \mathbf{n}$.

Rearrange columns of $M$ to get another $n \times n$ matrix $\xoverline[0.55]{M}$ such that for each $i \in \mathbf{n}$, the $i$th column of $M$ is the $f_{\pi}(f_0(i))$th column of $\xoverline[0.55]{M}$. Let $z_1$ be a valid term in $\det M$, which corresponds to a multi-colored subgraph $\mathbb{S}_1$ in $\mathcal{E}$. Let $\sigma_1 \in \mathcal{S}_n$ be the permutation associated with term $z_1$. That is, each parameter in $z_1$ is taken from a location in the $i$th row and the $\sigma_1(i)$th column of $M$ for some $i \in \mathbf{n}$. So the sign of $z_1$ is $\sgn(\sigma_1)$. Term $z_1$ naturally pairs with a valid term in $\det \xoverline[0.55]{M}$, denoted by $\bar{z}_1$. To be precise, if a parameter in $z_1$ is taken from the location in the $i$th row and the $\sigma_1(i)$th column of $M$, $\bar{z}_1$ has the same parameter taken from the location in the $i$th row and the $f_{\pi}(f_0(\sigma_1(i)))$th column of $\xoverline[0.55]{M}$. Let $\bar{\sigma}_1 \in \mathcal{S}_n$ be the permutation associated with term $\bar{z}_1$. So $\bar{\sigma}_1 = f_{\pi} f_0 \sigma_1$ and the sign of $\bar{z}_1$ is $\sgn(\bar{\sigma}_1)$. As $f_{\pi} f_0 \in \mathcal{S}_n$, 
\begin{equation*}
\sgn(\bar{\sigma}_1) = \sgn(f_{\pi} f_0) ~\sgn(\sigma_1) 
\end{equation*}

Let $c_1$ be the number of cycle graphs in the quotient graph $\mathbb{S}_1/\pi$. Let $\mathbb{Q}_1$ be the subgraph of $\mathbb{S}_1/\pi$ obtained by removing all the isolated vertices, if any, from $\mathbb{S}_1/\pi$. So $\mathbb{Q}_1$ is the disjoint union of $c_1$ cycle graphs. It can be checked that $\mathbb{Q}_1$ has $n$ vertices and $n$ arcs in $n$ distinct colors. In fact, $\mathbb{Q}_1$ is exactly the spanning subgraph of $\mathbb{G}_{\scriptsize{\xoverline[0.55]{M}}}$ that term $\bar{z}_1$ in $\det \xoverline[0.55]{M}$ corresponds to. By Proposition~\ref{prp:sgn-in-det}, $\sgn(\bar{\sigma}_1) = 1$ if $n-c_1$ is even, and $\sgn(\bar{\sigma}_1) = -1$ if $n-c_1$ is odd. It means that $c_1$ is even if $\sgn(\bar{\sigma}_1) = (-1)^n$, and $c_1$ is odd if $\sgn(\bar{\sigma}_1) = -(-1)^n$. So $\mathbb{S}_1$ is even if 
\begin{equation*}
\sgn(f_{\pi} f_0) ~\sgn(\sigma_1) = (-1)^n
\end{equation*}
and $\mathbb{S}_1$ is odd if 
\begin{equation*}
\sgn(f_{\pi} f_0) ~\sgn(\sigma_1) = -(-1)^n 
\end{equation*}
Let $z_2$ be another valid term in $\det M$, which corresponds to a multi-colored subgraph $\mathbb{S}_2$ in $\mathcal{E}$. Let $\sigma_2 \in \mathcal{S}_n$ be the permutation associated with term $z_2$. So the sign of $z_2$ is $\sgn(\sigma_2)$. Similarly, $\mathbb{S}_2$ is even if 
\begin{equation*}
\sgn(f_{\pi} f_0) ~\sgn(\sigma_2) = (-1)^n
\end{equation*}
and $\mathbb{S}_2$ is odd if 
\begin{equation*}
\sgn(f_{\pi} f_0) ~\sgn(\sigma_2) = -(-1)^n 
\end{equation*}
Therefore, the relative parity of $\mathbb{S}_1$ and $\mathbb{S}_2$ in $\mathcal{E}$ only depends on the relative sign of $z_1$ and $z_2$. If the two valid terms have the same sign, their corresponding multi-colored subgraphs have the same parity, and vice versa. \hfill $\qed$

\begin{lemma} \label{lem:genrnk-graph}
For a linearly parameterized matrix pair $(A, B)$ which satisfies the binary assumption, 
\begin{equation*}
\g-rank [A \enspace B] = n
\end{equation*} 
if and only if the graph of $(A, B)$ has an unbalanced similarity class of multi-colored subgraphs.
\end{lemma}

\noindent \textbf{Proof of Lemma~\ref{lem:genrnk-graph}:} When the binary assumption holds, $\g-rank [A \enspace B] = n$ if and only if there exists an $n \times n$ submatrix of $[A \enspace B]$, written $M$, such that $\g-rank M =  n$. By (\ref{eqn:det-bin-prm}), $\g-rank M = n$ if and only if $\exists \, \mathcal{C} \subset \mathbf{q}$, $|\mathcal{C}| = n$ such that $a_{\mathcal{C}} \neq 0$. As each valid term in $\det M$ is a signed product of $n$ distinct parameters, $a_{\mathcal{C}} \neq 0$ if and only if the number of positive valid terms $\prod_{k \in \mathcal{C}} p_k$ and the number of negative valid terms $-\prod_{k \in \mathcal{C}} p_k$ are not equal. By the proof of Lemma~\ref{lem:bijctn-parity}, a positive valid term $\prod_{k \in \mathcal{C}} p_k$ and a negative valid term $-\prod_{k \in \mathcal{C}} p_k$ correspond to two multi-colored subgraphs with opposite parities in the same similarity class. So $a_{\mathcal{C}} \neq 0$ if and only if the similarity class is unbalanced. Therefore, $\g-rank [A \enspace B] = n$ if and only if the graph of $(A, B)$ has an unbalanced similarity class of multi-colored subgraphs. \hfill $\qed$

\noindent \textbf{Proof of Theorem~\ref{thm:bin-prm-cond}, (\lowercase\expandafter{\romannumeral2})$\iff$(\lowercase\expandafter{\romannumeral4}):} By Lemma~\ref{lem:genrnk-graph} and Proposition~\ref{prp:irrdc-forest}, (\lowercase\expandafter{\romannumeral2})$\iff$(\lowercase\expandafter{\romannumeral4}). \hfill $\qed$

\subsection{Proof of Theorem~\ref{thm:bin-prm-cond}, (\lowercase\expandafter{\romannumeral3})$\iff$(\lowercase\expandafter{\romannumeral4})}

The following lemma makes the proof of the equivalence of statements (\lowercase\expandafter{\romannumeral3}) and (\lowercase\expandafter{\romannumeral4}) fairly straightforward.

\begin{lemma} \label{lem:forest-cactus}
Let $\mathbb{G}$ be a directed graph on $n+m$ vertices. Then $\mathbb{G}$ has a spanning subgraph which is a disjoint union of $m$ cactus graphs rooted at $m$ distinct vertices if and only if $\mathbb{G}$ has two spanning subgraphs: One is a spanning forest rooted at the same $m$ vertices; The other is a disjoint union of $m$ path graphs and a non-negative number of cycle graphs, where the source vertices are the $m$ root vertices of the cactus graphs.
\end{lemma}

Note that Lemma~\ref{lem:forest-cactus} has no requirement on the color of arcs.

\noindent \textbf{Proof of Lemma~\ref{lem:forest-cactus}:} The necessity is obvious. Let us prove the sufficiency. Let $\mathbb{U}$ be a spanning subgraph of $\mathbb{G}$, which is the disjoint union of $m$ path graphs and $c$ cycle graphs. If $c=0$, $\mathbb{U}$ is already a disjoint union of $m$ cactus graphs with no bud. Now assume $c>0$.

Let $\mathcal{V}$ be the vertex set of $\mathbb{G}$. Let $\mathcal{V}_0 \subset \mathcal{V}$ be the set of vertices in the $m$ path graphs of $\mathbb{U}$. Let $\mathcal{V}_{\text{root}} \subset \mathcal{V}_0$ be the set of source vertices of the $m$ path graphs in $\mathbb{U}$. For each $i \in \mathbf{c} \triangleq \{1, 2, \dots, c\}$, let $\mathcal{V}_i \subset \mathcal{V}$ be the set of vertices in the $i$th cycle graph of $\mathbb{U}$. So 
\begin{equation*}
\mathcal{V} = \bigcup\limits_{i=0}^c \mathcal{V}_i
\end{equation*}
and 
\begin{equation*}
\mathcal{V}_i \cap \mathcal{V}_j = \emptyset, ~i \neq j, ~i, j \in \{0, 1, \dots, c\} 
\end{equation*}
Since $\mathbb{G}$ has a spanning forest rooted at the $m$ vertices in $\mathcal{V}_{\text{root}}$, there exists an arc in $\mathbb{G}$ from a vertex in $\mathcal{V}_0$ to a vertex in $\mathcal{V}_{i_1}$ for some $i_1 \in \mathbf{c}$. Otherwise, there is no path to the vertices in $\mathcal{V}-\mathcal{V}_0$ from any root vertex. Let $\mathcal{V}_0^1 \triangleq \mathcal{V}_0 \cup \mathcal{V}_{i_1}$. Similarly, there exists another arc in $\mathbb{G}$ from a vertex in $\mathcal{V}_0^1$ to a vertex in $\mathcal{V}_{i_2}$ for some $i_2 \in \mathbf{c} - \{i_1\}$, otherwise there is no path to the vertices in $\mathcal{V}-\mathcal{V}_0^1$ from any root vertex. The process continues until one finds $c$ arcs in $\mathbb{G}$ that connect $\mathcal{V}_0$, $\mathcal{V}_1$, $\dots$, $\mathcal{V}_c$. The addition of the $c$ arcs to $\mathbb{U}$ renders a disjoint union of $m$ cactus graphs rooted at the $m$ vertices in $\mathcal{V}_{\text{root}}$. \hfill $\qed$

\noindent \textbf{Proof of Theorem~\ref{thm:bin-prm-cond}, (\lowercase\expandafter{\romannumeral3})$\iff$(\lowercase\expandafter{\romannumeral4}):} Obviously, (\lowercase\expandafter{\romannumeral3}) $\Longrightarrow$ (\lowercase\expandafter{\romannumeral4}). If the graph $\mathbb{G}$ of $(A, B)$ has an unbalanced similarity class of multi-colored subgraphs, $\mathbb{G}$ has at least one multi-colored subgraph. So a spanning subgraph of $\mathbb{G}$ is the disjoint union of $m$ path graphs and a non-negative number of cycle graphs, where the source vertices are the $m$ vertices with labels $n+1$ to $n+m$. By Lemma~\ref{lem:forest-cactus}, (\lowercase\expandafter{\romannumeral4}) $\Longrightarrow$ (\lowercase\expandafter{\romannumeral3}). Therefore, (\lowercase\expandafter{\romannumeral3})$\iff$(\lowercase\expandafter{\romannumeral4}). \hfill $\qed$

\section{Conclusion}

This paper extends the graph-theoretic conditions for structural controllability to the class of linearly parameterized matrix pairs satisfying the binary assumption. As a byproduct of the analysis, Corollary~\ref{crl:rankAB} presents a tighter upper bound on the rank of a matrix product than the minimum rank of the matrices in the product. If one wants to further extend the graph-theoretic conditions to all linearly parameterized matrix pairs, weighted graphs of matrix pairs must be introduced. To accommodate this, some graphical concepts will have to be modified accordingly, such as quotient graph, multi-colored subgraph, balanced or unbalanced similarity class of multi-colored subgraphs, and line graph. Some future research problems are: (1) to show that it is NP-hard to determine whether the graph of $(A, B)$ has an unbalanced similarity class of multi-colored subgraphs; (2) to find the minimum number of input required for the structural controllability of a given linearly parameterized matrix $A_{n \times n}$; (3) to study the structural controllability of linearly parameterized linear time-varying systems; (4) to eventually generalize the definition and the corresponding characterizations of structural controllability to nonlinear systems for which there is a good understanding of controllability.


\bibliographystyle{IEEEtran}
\bibliography{IEEEabrv,fjbibTAC}



%
%
%
%
%
%


\end{document}